\documentstyle[aps,prb,twocolumn,epsf,floats]{revtex}
\draft
\begin{document}
\wideabs{

\title{Energy, interaction, and photoluminescence 
       of spin-reversed quasielectrons\\
       in fractional quantum Hall systems}

\author{
   Izabela Szlufarska}
\address{
   Department of Physics,
   University of Tennessee, Knoxville, Tennessee 37996, \\
   and Institute of Physics, 
   Wroclaw University of Technology, Wroclaw 50-370, Poland}

\author{
   Arkadiusz W\'ojs}
\address{
   Department of Physics, 
   University of Tennessee, Knoxville, Tennessee 37996, \\
   and Institute of Physics, 
   Wroclaw University of Technology, Wroclaw 50-370, Poland}

\author{
   John J. Quinn}
\address{
   Department of Physics, 
   University of Tennessee, Knoxville, Tennessee 37996}

\maketitle

\begin{abstract}
The energy and photoluminescence spectra of a two-dimensional 
electron gas in the fractional quantum Hall regime are studied.
The single-particle properties of reversed-spin quasielectrons
(QE$_{\rm R}$'s) as well as the pseudopotentials of their 
interaction with one another and with Laughlin quasielectrons
(QE's) and quasiholes (QH's) are calculated.
Based on the short-range character of the QE$_{\rm R}$--QE$_{\rm R}$ 
and QE$_{\rm R}$--QE repulsion, the partially unpolarized 
incompressible states at the filling factors $\nu={4\over11}$ 
and ${5\over13}$ are postulated within Haldane's hierarchy scheme.
To describe photoluminescence, the family of bound $h($QE$_{\rm R})_n$ 
states of a valence hole $h$ and $n$ QE$_{\rm R}$'s are predicted 
in analogy to the found earlier fractionally charged excitons 
$h$QE$_n$.
The binding energy and optical selection rules for both families
are compared.
The $h$QE$_{\rm R}$ is found radiative in contrast to the dark 
$h$QE, and the $h($QE$_{\rm R})_2$ is found non-radiative in 
contrast to the bright $h$QE$_2$.
\end{abstract}
}

\section{Introduction}

The integer\cite{klitzing} and fractional\cite{tsui,laughlin1,prange} 
quantum Hall effects (IQHE and FQHE) both depend on the finite gap 
$\Delta$ for charge excitations that opens in a two-dimensional 
electron gas (2DEG) at the specific (integral or fractional) filling 
factors $\nu$, defined as the number of electrons $N$ divided by the 
Landau level (LL) degeneracy $g$.
At sufficiently low temperatures, this gap makes the system 
incompressible and, among other effects, forbids electric 
conductance and causes quantization of Hall resistivity.

It is quite remarkable that the most prominent FQH states, so-called 
Laughlin ground states\cite{laughlin1} that occur at $\nu=(2p+1)^{-1}$ 
($p$ is an integer), are the only ones that are maximally spin-polarized 
solely due to the electron--electron exchange interaction.
At other filling factors, the 2DEG is known\cite{halperin1,chakraborty2,%
zhang,xie,wu1,chakraborty1,apalkov} to be at least partially unpolarized 
unless the Zeeman energy $E_{\rm Z}$ is sufficiently large.
Only partial polarization of the FQH states at the filling factors
other than $\nu=(2p+1)^{-1}$ causes transitions\cite{chakraborty3} 
between incompressible and compressible or different incompressible 
phases as a function of $E_{\rm Z}$, realized in tilted-field 
experiments.\cite{clark,eisenstein,engel,willet} 
The finite excitation gap $\Delta$ of the Laughlin state results from 
the finite energies $\varepsilon$ of its elementary charge excitations, 
Laughlin quasielectrons (QE's) and quasiholes (QH's), as well as from 
the lack of the particle--hole symmetry between them that causes 
a magneto-roton type of dispersion of the QE--QH interaction with 
a minimum at a finite wave vector $k$.
Indeed, the calculated\cite{haldane1,laughlin2,haldane2,haldane3,fano,%
wojs-pmb} energy $\varepsilon_{\rm QE}+\varepsilon_{\rm QH}$ needed to 
create a spatially separated QE--QH pair necessary for electric current 
agreed reasonably well with the activation energy obtained from the 
temperature dependences of the FQHE at $\nu=(2p+1)^{-1}$.

Therefore, it was quite surprising when Rezayi\cite{rezayi} and
Chakraborty et al.\cite{chakraborty4} discovered that another 
low-energy excitation of the Laughlin state exists, a spin-density 
wave, which becomes gapless at $E_{\rm Z}=0$.
It turns out that it is only due to a finite Zeeman energy that the 
spontaneous creation of spin waves, each consisting of a positively 
charged QH and a negatively charged reversed-spin quasielectron 
(QE$_{\rm R}$), does not destroy incompressibility of Laughlin states 
in the experimental 2DEG systems.
Although the spin excitations of Laughlin states have been extensively 
studied in the context of the real-space spin patterns called skyrmions
\cite{lee,sondhi} (particularly at $\nu=1$), our knowledge of their 
interaction with one another or with other excitations, or their 
optical properties is not yet complete (specially at fractional $\nu$).
In this paper we address both of these issues.

First, we identify QE, QH, and QE$_{\rm R}$ as the three elementary 
quasiparticles (QP's) of a Laughlin state and determine their mutual 
interaction pseudopotentials $V({\cal R})$, defined\cite{haldane3} as 
the dependence of the pair interaction energy $V$ on the relative pair 
angular momentum ${\cal R}$.
For example, the QE$_{\rm R}$--QE$_{\rm R}$ pseudopotential is found 
to be very different from the QE--QE pseudopotential at short range, 
which is the reason for incompressibility of a partially polarized 
$\nu={4\over11}$ state at low $E_{\rm Z}$ (in contrast to the 
compressible\cite{wojs-hierarchy} fully polarized state at the 
same $\nu$).
A partially polarized $\nu={4\over11}$ state has been also recently 
proposed by Park and Jain\cite{park} within a composite fermion
\cite{jain,lopez,halperin2} (CF) model.
However, their interpretation of the $\nu={4\over11}$ as a mixed 
state of CF's with two and four attached vortices (fluxes) is not 
very accurate in a sense that the two additional vortices (fluxes) 
attached to each spin-reversed CF are not vortices of the many-body 
wave function expressed in terms of the same coordinates (fluxes 
of the same effective magnetic field) as the original two attached 
to each electron (to form CF's).
The correct interpretation necessarily involves reapplication of the 
CF transformation to some of the original CF's (those in a partially 
filled reversed-spin LL), in analogy to the CF hierarchy proposed 
by Sitko et al.\cite{sitko,quinn} and essentially equivalent
\cite{wojs-hierarchy} to Haldane's hierarchy.\cite{haldane1}
Let us stress that it is the short range of the QE$_{\rm R}$--QE$_{\rm R}$ 
repulsion shown here that justifies application of Haldane hierarchy 
to QE$_{\rm R}$'s (or, equivalently, spin-reversed CF's).

Second, in analogy to the fractionally charged excitons (FCX's)
\cite{chen,wojs-fcx} consisting of a number of QE's of a 
spin-polarized 2DEG bound to a valence-band hole $h$, we discuss the
possible formation and radiative recombination of similar complexes 
denoted as FCX$_{\rm R}$'s and containing one or more QE$_{\rm R}$'s 
bound to a hole.
We find that different optical selection rules for FCX's and 
FCX$_{\rm R}$'s could allow optical detection of QE$_{\rm R}$'s 
in the 2DEG without need for direct polarization measurement.

\section{Model}

The properties of spin-reversed quasielectrons (QE$_{\rm R}$) are 
studied by exact numerical diagonalization in an ideal 2DEG with 
zero width and no disorder.
The magnetic field $B$ is assumed to be sufficiently large 
(the cyclotron energy $\hbar\omega_c\propto B$ much larger than 
the interaction energy scale $e^2/\lambda\propto\sqrt{B}$, where 
$\lambda$ is the magnetic length) that only the lowest LL need 
be considered.
In order to describe an infinite planar system with 2D translational 
symmetry in a finite-size calculation we use Haldane's spherical 
geometry\cite{haldane1} in which the (finite) LL degeneracy $g=2S+1$ 
is controlled by the strength $2S$ of the magnetic monopole placed 
in the center of the sphere of radius $R$.
The monopole strength $2S$ is defined in the units of flux quantum 
$\phi_0=hc/e$, so that $4\pi R^2B=2S\phi_0$ and $R^2=S\lambda^2$.
The single-particle states on a sphere $\left|S,l,m\right>$ are called 
monopole harmonics.\cite{haldane1,fano,wu2}
They are eigenstates of length $l$ and projection $m$ of angular 
momentum and form LL's labeled by $n=l-S$, analogous to those of 
planar geometry.
The lowest LL included in the present calculation has $n=0$ and $l=S$,
and its orbitals are simply denoted by $\left|m\right>$ with $|m|\le S$.
The electronic spin is included in the model by adding a quantum number 
$\sigma$ denoting the projection of spin.
As usual, the Zeeman term is taken as $E_{\rm Z}\propto B\sigma$ to 
avoid an unphysical spin-orbit coupling resulting for $E_{\rm Z}\propto$ 
{\boldmath$B\sigma$} and for a heterogeneous (radial) magnetic field 
on a sphere.

The many-electron interaction Hamiltonian reads
\begin{equation}
   H_{ee}=
   \sum
   c_{m_1\sigma}^\dagger
   c_{m_2\sigma'}^\dagger
   c_{m_3\sigma'}
   c_{m_4\sigma}
   \left<m_1m_2|V_{ee}|m_3m_4\right>,
\label{eqhamee}
\end{equation}
where operators $c_{m\sigma}^\dagger$ and $c_{m\sigma}$ create
and annihilate an electron in the state $\left|m\sigma\right>$,
the summations go over all orbital and spin indices, and the two-body 
interaction matrix elements are calculated for the Coulomb potential 
$V_{ee}(r)=e^2/r$.
Hamiltonian $H_{ee}$ is diagonalized in the basis of $N$-electron 
Slater determinants
\begin{equation}
   \left|m_1\sigma_1 \dots m_N\sigma_N\right>=
   c_{m_1\sigma_1}^\dagger 
   \dots 
   c_{m_N\sigma_N}^\dagger
   \left|{\rm vac}\right>,
\label{eqbasee}
\end{equation}
where $\left|{\rm vac}\right>$ stands for the vacuum state.
While using basis (\ref{eqbasee}) allows automatic resolution of 
two good many-body quantum numbers, projection of spin ($J_z=
\sum\sigma_i$) and angular momentum ($L_z=\sum m_i$), the other 
two, length of spin ($J$) and angular momentum ($L$), are resolved 
numerically in the diagonalization of each appropriate $(J_z,L_z)$ 
Hilbert subspace.

In order to describe the reversed-spin fractionally charged exciton 
(FCX$_{\rm R}$) states, a single valence-band hole $h$ is added 
to the model $N$-electron system.
Since, as for FCX's, the formation of FCX$_{\rm R}$ states requires 
weakening of the electron--hole attraction compared to the 
electron--electron repulsion,\cite{wojs-fcx} the hole is placed 
on a parallel plane, separated by a distance $d$ (of the order of 
$\lambda$) from the 2DEG.
Because the physics of an isolated FCX or FCX$_{\rm R}$ to a good 
approximation does not depend on the (possibly complicated) 
structure of the valence band, the single-hole wave functions are 
taken the same as for electrons (except for the reversed signs of 
$m$ and $\sigma$).
This means that both inter-LL hole scattering and the mixing 
between heavy- and light-hole subbands are ignored.
The weak electron--hole exchange is also neglected so that the hole 
spin has no effect on the dynamics of an FCX or FCX$_{\rm R}$, and 
the interaction of a hole with the 2DEG is described by the following 
spin-conserving term
\begin{equation}
   H_{eh}=
   \sum
   c_{m_1\sigma}^\dagger
   h_{m_2}^\dagger
   h_{m_3}
   c_{m_4\sigma}
   \left<m_1m_2|V_{eh}|m_3m_4\right>
\label{eqhameh}
\end{equation}
in the total Hamiltonian $H=H_{ee}+H_{eh}$.
In the above, operators $h_m^\dagger$ and $h_m$ create and annihilate
a hole in the orbital $\left|m\right>$ of the valence band, and the 
electron--hole interaction is defined by the Coulomb potential
$V_{eh}(r)=-e^2/\sqrt{r^2+d^2}$.
The exclusion of the hole--hole interaction effects from $H$ reflects 
the fact that $\nu_h\ll\nu$.
Interaction Hamiltonian $H$ is diagonalized in the basis of 
single-particle configurations
\begin{equation}
   \left|m_1\sigma_1 \dots m_N\sigma_N;m_h\right>=
   c_{m_1\sigma_1}^\dagger 
   \dots 
   c_{m_N\sigma_N}^\dagger
   h_{m_h}^\dagger
   \left|{\rm vac}\right>,
\label{eqbaseh}
\end{equation}
and the set of good quantum numbers labelling many-electron--one-hole 
eigenstates includes $J_z$ and $J$ of the electrons, hole spin $\sigma_h$ 
(omitted in our equations), and the length ($L$) and projection ($L_z$) 
of angular momentum of the total electron--hole system.

The justification for using Haldane's spherical geometry to model an 
infinite planar 2DEG (with or without additional valence holes) relies 
on the exact mapping between the orbital numbers $L$ and $L_z$ and the 
two good quantum numbers on a plane (resulting from the 2D translational
symmetry), angular momentum projection ${\cal M}$ and an additional
angular momentum quantum number ${\cal K}$ associated with partial 
decoupling of the center-of-mass motion of an electron--hole system 
in a homogeneous magnetic field.\cite{avron,dzyubenko}
This mapping guarantees correct description of such symmetry-dependent 
effects as degeneracies in the energy spectrum or the optical selection 
rules (associated with conservation of ${\cal M}$ and ${\cal K}$ or $L$ 
and $L_z$ in the absorption or emission of a photon).
The energy values obtained on a sphere generally depend on the 
surface curvature, that is on $R/\lambda=\sqrt{S}$.
However, for those energies that describe finite-size objects 
(such as QE$_{\rm R}$ or FCX$_{\rm R}$ studied here) or their 
interaction at a finite range (here, pseudopotential parameters for 
interaction of QE$_{\rm R}$ with other particles), the values 
characteristic of an infinite planar system can be estimated from 
the calculation done for sufficiently large $2S$ and $N$ (or 
extrapolation of finite-size data to the $2S\rightarrow\infty$ 
limit).

\section{Spin-Reversed Quasielectrons: Results and Discussion}

\subsection{Stability and Single-Particle Properties}

\begin{figure}[t]
\epsfxsize=3.40in
\epsffile{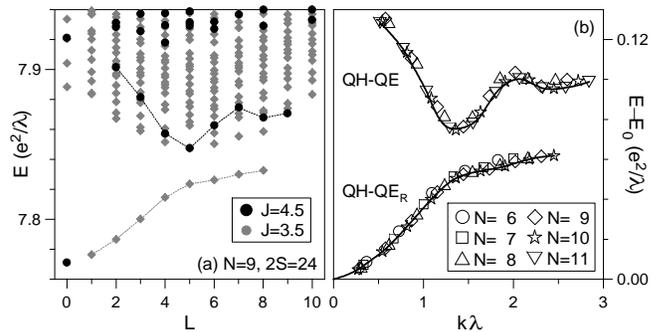}
\caption{
   (a) The energy spectrum (Coulomb energy $E$ versus angular 
   momentum $L$) of the system of $N=9$ electrons on Haldane 
   sphere at the monopole strength $2S=3(N-1)=24$.
   Black dots and grey diamonds mark states with the total 
   spin $J={1\over2}N={9\over2}$ (maximum polarization) 
   and $J={1\over2}N-1={7\over2}$ (one reversed spin), 
   respectively.
   Ground state is the Laughlin $\nu={1\over3}$ state.
   Lines connect states containing one QE--QH ($J={9\over2}$) 
   or QE$_{\rm R}$--QH ($J={7\over2}$) pair.
   (b) The dispersion curves (excitation energy ${\cal E}_\Sigma
   =E-E_0$ versus wave vector $k$) for the $\Sigma=0$ charge-density 
   wave (QE--QH pair) and the $\Sigma=1$ spin-density wave 
   (QE$_{\rm R}$--QH pair) in the Laughlin $\nu={1\over3}$ ground 
   state, calculated in the systems of $N\le11$ electrons on Haldane 
   sphere. 
   $\lambda$ is the magnetic length.}
\label{fig1}
\end{figure}

It is well-known that even in the absence of the Zeeman energy gap,
$E_{\rm Z}=0$, the ground state of the 2DEG in the lowest LL is 
completely spin-polarized at the precise values of the Laughlin filling 
factor $\nu=(2p+1)^{-1}$, with $p=0$, 1, 2, \dots.
There are two types of elementary charge-neutral excitations of 
Laughlin $\nu=(2p+1)^{-1}$ ground states, carrying spin $\Sigma=0$ 
or 1, respectively.
Their dispersion curves (energy as a function of wave vector), 
${\cal E}_\Sigma(k)$, have been studied for all combinations of $p$ 
and $\Sigma$.
While the formulae for the $\nu=1$ ground state have been evaluated 
analytically,\cite{gorkov,bychkov,kallin} in Fig.~\ref{fig1} we present 
the exact numerical results for $\nu={1\over3}$ obtained from our exact 
diagonalization of up to $N=11$ electrons on Haldane's sphere.
As an example, in Fig.~\ref{fig1}(a), we show the entire low-energy 
spectrum of an $N=9$ system with all spins polarized and with one 
reversed spin (Hilbert subspaces of total spin $J={1\over2}N-\Sigma=
{9\over2}$ and ${7\over2}$ for $\Sigma=0$ and 1, respectively), from 
which the dispersion curves ${\cal E}_\Sigma(k)$ are obtained.
The energy $E$ is plotted as a function of angular momentum $L$, and 
$2S=3(N-1)=24$ is the strength of the magnetic monopole inside Haldane's 
sphere corresponding to the LL degeneracy $g=2S+1=25$ and the Laughlin 
filling factor $\nu=(N-1)/(g-1)={1\over3}$ (for the details of Haldane's 
spherical geometry see Refs.~\onlinecite{haldane1,fano,wu2}).
The energy $E$ does not include the Zeeman term $E_{\rm Z}$, which scales 
differently than the plotted Coulomb energy with the magnetic field $B$.
The excitation energies ${\cal E}_\Sigma=E-E_0$ (where $E_0$ is the 
Laughlin ground state energy) have been calculated for the states 
identified in the finite-size spectra as the $\Sigma=0$ charge-density 
wave and the $\Sigma=1$ spin-density wave.
These states are marked with dotted lines in Fig.~\ref{fig1}(a).
The values of ${\cal E}_\Sigma$ obtained for different $N\le11$ have 
been plotted together in Fig.~\ref{fig1}(b) as a function of the wave 
vector $k=L/R=(L/\sqrt{S})\lambda^{-1}$.
Clearly, using the appropriate units of $\lambda^{-1}$ for wave 
vector and $e^2/\lambda$ for excitation energy in Fig.~\ref{fig1}(b) 
results in the quick convergence of the curves with increasing $N$, 
and allows accurate prediction of the dispersion curves in an infinite 
system, as marked with thick lines.
The most significant features of these curves are: 
(i) the finite gap $\Delta_0\approx0.076\,e^2/\lambda$ and the 
magneto-roton minimum $k\approx1.5\lambda^{-1}$ in ${\cal E}_0(k)$, and
(ii) the vanishing of ${\cal E}_1(k)$ in the $k\rightarrow0$ limit 
(for $E_{\rm Z}=0$).

The similar nature of the charge- and spin-waves in the $\nu={1\over3}$ 
state to those at $\nu=1$ lies at the heart of the composite fermion 
(CF) picture,\cite{jain,lopez,halperin2} in which these excitations 
correspond to promoting one CF from a completely filled lowest ($n=0$) 
spin-$\downarrow$ CF LL either to the first excited ($n=1$) CF LL of 
the same spin ($\downarrow$) or to the same CF LL ($n=0$) but with 
the reversed spin ($\uparrow$).
The three constituent QP's from which the charge- and spin-waves are 
composed: a hole in the $n=0$ spin-$\downarrow$ CF LL and particles 
in the $n=1$ spin-$\downarrow$ and $n=0$ spin-$\uparrow$ CF LL's, are 
analogous to those in the electron LL's from which the charge- and 
spin-waves at $\nu=1$ are built.

Independently of the CF picture, one can define three types of QP's 
(elementary excitations) of the Laughlin $\nu={1\over3}$ fluid.
They are Laughlin quasiholes (QH's) and quasielectrons (QE's) and 
Rezayi spin-reversed quasielectrons (QE$_{\rm R}$).
The excitations in Fig.~\ref{fig1} are more complex in a sense that 
they consist of a (neutral) pair of QH and either QE ($\Sigma=0$) or 
QE$_{\rm R}$ ($\Sigma=1$).
Each of the QP's is characterized by such single-particle quantities 
as (fractional) electric charge (${\cal Q}_{\rm QH}=+{1\over3}e$ 
and ${\cal Q}_{\rm QE}={\cal Q}_{\rm QER}=-{1\over3}e$), energy 
$\varepsilon_{\rm QP}$, or degeneracy $g_{\rm QP}$ of the 
single-particle Hilbert space.
On Haldane's sphere, the degeneracy $g_{\rm QP}$ is related to the 
angular momentum $l_{\rm QP}$ by $g_{\rm QP}=2l_{\rm QP}+1$, with 
$l_{\rm QH}=l_{\rm QER}=S^*$ and $l_{\rm QE}=S^*-1$ and $2S^*=2S-2(N-1)$ 
being the effective monopole strength in the CF model.

\begin{figure}[t]
\epsfxsize=3.40in
\epsffile{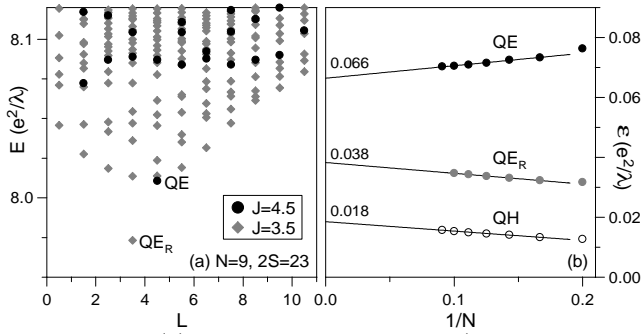}
\caption{
   (a) The energy spectrum (Coulomb energy $E$ versus angular 
   momentum $L$) of the system of $N=9$ electrons on Haldane 
   sphere at the monopole strength $2S=3(N-1)-1=23$.
   Black dots and grey diamonds mark states with the total 
   spin $J={1\over2}N={9\over2}$ (maximum polarization) 
   and $J={1\over2}N-1={7\over2}$ (one reversed spin), 
   respectively.
   Ground state at $J={7\over2}$ is the reversed-spin 
   quasielectron QE$_{\rm R}$ of the Laughlin $\nu={1\over3}$ 
   fluid, and the lowest-energy state at $J={9\over2}$ is the 
   Laughlin quasielectron QE.
   (b) The energies $\varepsilon$ of all three types of 
   quasiparticles of the Laughlin $\nu={1\over3}$ ground state
   (QH, QE, and QE$_{\rm R}$) calculated in the systems of $N\le11$ 
   electrons on Haldane sphere and plotted as a function of $N^{-1}$.
   The numbers give the results of linear extrapolation to an
   infinite (planar) system.
   $\lambda$ is the magnetic length.}
\label{fig2}
\end{figure}

The energies $\varepsilon_{\rm QP}$ to create an isolated QP 
of each type in the Laughlin ground state have been previously 
estimated in a number of ways.
Here, we present our results of exact diagonalization calculation 
for $N\le11$ ($\varepsilon_{\rm QE}$ and $\varepsilon_{\rm QH}$) 
and $N\le10$ ($\varepsilon_{\rm QER}$).
In Fig.~\ref{fig2}(a) we show an example of the numerical energy 
spectrum for the system of $N=9$ electrons, in which an isolated 
QE or QE$_{\rm R}$ occurs at $2S=3(N-1)-1=23$ in the subspace of 
$J={1\over2}N={9\over2}$ and $J={1\over2}N-1={7\over2}$, respectively.
Both of these states have been identified in Fig.~\ref{fig2}(a).
To estimate $\varepsilon_{\rm QE}$ and $\varepsilon_{\rm QER}$, 
we use the standard procedure\cite{haldane2,haldane3,fano,wojs-pmb,%
wojs-hierarchy} to account for the finite-size effects (dependence of 
$\lambda$ on $2S$, $S\lambda^2=R^2$), and express the energies $E$ of 
Fig.~\ref{fig2}(a) in the units of $e^2/\lambda$ with $\lambda$ 
appropriate for $\nu={1\over3}$, before subtracting from them the 
Laughlin ground state energy of Fig.~\ref{fig1}(a).
Plotting the results for different values of $N$ in Fig.~\ref{fig2}(b) 
as a function of $N^{-1}$ allows the extrapolation to an infinite system, 
with the limiting values of $\varepsilon_{\rm QE}=0.0664\,e^2/\lambda$ 
and $\varepsilon_{\rm QER}=0.0383\,e^2/\lambda$ (with the difference
$\varepsilon_{\rm QE}-\varepsilon_{\rm QER}=0.0281\,e^2/\lambda$ 
in remarkable agreement with Rezayi's original estimate\cite{rezayi} 
based on his numerics for $N\le6$).
For completeness, we have also plotted the QH energies, which 
extrapolate to $\varepsilon_{\rm QH}=0.0185\,e^2/\lambda$.
Note that to obtain the so-called ``proper'' QP energies in a finite
system,\cite{haldane2,fano,wojs-pmb} $\tilde{\varepsilon}_{\rm QP}(N)$, 
the term ${\cal Q}_{\rm QP}^2/2R$ must be added to each value in 
Fig.~\ref{fig2}(b).
The linear extrapolation of $\tilde{\varepsilon}_{\rm QP}(N)$ to 
$N^{-1}\rightarrow0$ gives 
$\tilde{\varepsilon}_{\rm QE}=0.0737\,e^2/\lambda$,
$\tilde{\varepsilon}_{\rm QER}=0.0457\,e^2/\lambda$, and 
$\tilde{\varepsilon}_{\rm QH}=0.0258\,e^2/\lambda$.
The energies of spatially separated QE--QH and QE$_{\rm R}$--QH pairs 
(activation energies in transport experiments) are hence equal to
${\cal E}_0(\infty)=\tilde{\varepsilon}_{\rm QE}+
\tilde{\varepsilon}_{\rm QH}=0.0995\,e^2/\lambda$ and 
${\cal E}_1(\infty)=\tilde{\varepsilon}_{\rm QER}+
\tilde{\varepsilon}_{\rm QH}=0.0715\,e^2/\lambda$.

While the QH's are the only type of QP's that occur in low-energy 
states at $\nu<(2p+1)^{-1}$, the QE's and QE$_{\rm R}$'s are two 
competing excitations at $\nu>(2p+1)^{-1}$.
As pointed out by Rezayi\cite{rezayi} and Chakraborty et al.,
\cite{chakraborty4} whether QE's or QE$_{\rm R}$'s will occur at 
low energy depends on the relation between their energies including 
the Zeeman term, $\varepsilon_{\rm QE}$ and $\varepsilon_{\rm QER}+
E_{\rm Z}$.
Although it is difficult to accurately estimate the value of $E_{\rm Z}$ 
in an experimental sample because of its dependence on a number of 
factors (material parameters, well width $w$, density $\varrho$, 
magnetic field $B$, etc.), it seems that both scenarios with QE's 
and QE$_{\rm R}$'s being lowest-energy QP's are possible.
For example, using the bulk value for the effective $g^*$-factor in 
GaAs ($dE_{\rm Z}/dB=0.03$~meV/T) results in the QE$_{\rm R}$--QE crossing 
at $B=18$~T, while including the dependence of $g^*$ on $w$ and $B$ 
as described in Ref.~\onlinecite{wojs-xminus} makes QE$_{\rm R}$ more 
stable than QE up to $B\sim100$~T.

\subsection{Interaction with Other Quasiparticles}

Once it is established which of the QP's occur at low energy in 
a particular system (defined by $\varrho$, $w$, $B$, $\nu$, etc.), 
their correlations can be understood by studying the appropriate pair 
interaction pseudopotentials.\cite{haldane3,wojs-pmb,quinn,wojs-five}
The pseudopotential $V({\cal R})$ is defined\cite{haldane3} as the 
dependence of pair interaction energy $V$ on relative orbital angular 
momentum ${\cal R}$.
On a plane, ${\cal R}$ for a pair of particles $ab$ is the angular 
momentum associated with the (complex) relative coordinate, $z=z_a-z_b$.
On Haldane's sphere, the compatible definition of ${\cal R}$ depends
on the sign of ${\cal Q}_a{\cal Q}_b$:
for a pair of opposite charges, ${\cal R}$ is the length of total pair 
angular momentum, $L=|{\bf l}_a+{\bf l}_b|$, while for two charges of 
the same sign, ${\cal R}=l_a+l_b-L$.
In all cases, ${\cal R}\ge0$ and larger ${\cal R}$ corresponds to 
a larger average $ab$ separation.\cite{wojs-pmb,quinn}
Furthermore, only odd values of ${\cal R}$ are allowed for 
indistinguishable ($a=b$) fermions.

\begin{figure}[t]
\epsfxsize=3.40in
\epsffile{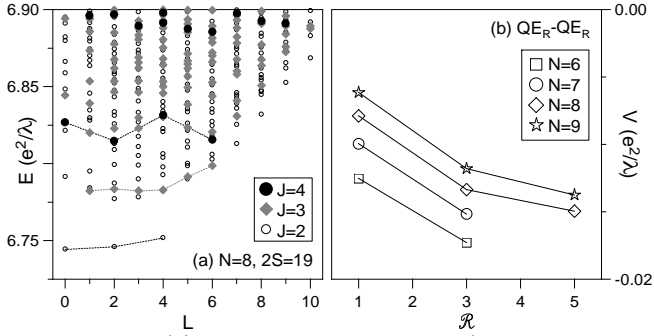}
\caption{
   (a) The energy spectrum (Coulomb energy $E$ versus angular 
   momentum $L$) of the system of $N=8$ electrons on Haldane 
   sphere at the monopole strength $2S=3(N-1)-2=19$.
   Black dots, grey diamonds, and open circles mark states with 
   the total spin $J={1\over2}N=4$ (maximum polarization), 
   $J={1\over2}N-1=3$ (one reversed spin), and $J={1\over2}N-2=2$ 
   (two reversed spins), respectively.
   Lines connect states containing one QE--QE ($J=4$),
   QE--QE$_{\rm R}$ ($J=3$), or QE$_{\rm R}$--QE$_{\rm R}$ 
   ($J=2$) pair.
   (b) The pseudopotentials (pair energy $V$ versus relative angular 
   momentum ${\cal R}$) of the QE$_{\rm R}$--QE$_{\rm R}$ interaction
   calculated in the systems of $N\le9$ electrons on Haldane sphere.
   $\lambda$ is the magnetic length.}
\label{fig3}
\end{figure}

Since the QE--QH and QE$_{\rm R}$--QH pseudopotentials have been plotted 
in Fig.~\ref{fig1} ($V_{\rm QE-QH}={\cal E}_0$ and $V_{\rm QER-QH}={\cal 
E}_1$), and the QE--QE and QH--QH pseudopotentials can be found for 
example in Ref.~\onlinecite{wojs-hierarchy}, we only need to discuss 
$V_{\rm QER-QER}$ and $V_{\rm QE-QER}$.
Two QE$_{\rm R}$'s occur in an $N$-electron system with at least two 
reversed spins ($J\le{1\over2}N-2$) and at $2S=3(N-1)-2$ (i.e., at 
$g=g_0-2$ where $g_0$ corresponds to the Laughlin state).
An example of the energy spectrum is shown in Fig.~\ref{fig3}(a) for 
$N=8$ at $2S=19$.
The lowest-energy states in the subspaces of $J={1\over2}N=4$, 
${1\over2}N-1=3$, and ${1\over2}N-2=2$ are connected with dashed lines
and contain a QE--QE, QE--QE$_{\rm R}$, and QE$_{\rm R}$--QE$_{\rm R}$
pair, respectively.
The angular momenta $L$ that occur in these bands result from addition
of ${\bf l}_{\rm QE}$ and/or ${\bf l}_{\rm QER}$ (with $l_{\rm QE}=S^*+1
={7\over2}$ and $l_{\rm QER}=S^*={5\over2}$).
For identical fermions, the addition must be followed by 
antisymmetrization that picks out only odd values of ${\cal R}$ for the 
QE--QE and QE$_{\rm R}$--QE$_{\rm R}$ pairs.
 
An immediate conclusion from Fig.~\ref{fig3}(a) is that the maximally 
spin-polarized ($J={1\over2}N$) system is unstable at the filling factor 
close but not equal to the Laughlin value of $\nu={1\over3}$ (the actual 
spin polarization decreases with decreasing $E_{\rm Z}$, and $J=0$ for 
$E_{\rm Z}=0$).
This was first pointed out by Rezayi\cite{rezayi} and interpreted in 
terms of an effective attraction between $\Sigma=1$ spin-waves; in this 
paper we prefer to use charged QP's as the most elementary excitations 
and explain the observed ordering of different $J$-bands by the fact 
that $\varepsilon_{\rm QE}\ne\varepsilon_{\rm QER}$ (at $E_{\rm Z}=0$, 
$\varepsilon_{\rm QE}-\varepsilon_{\rm QER}\approx0.0281\,e^2/\lambda$) 
and the particular form of involved interaction pseudopotentials
(see further in the text).

We have calculated the QE--QE$_{\rm R}$ and QE$_{\rm R}$--QE$_{\rm R}$ 
pseudopotentials from the energy spectra as that in Fig.~\ref{fig3}(a) 
by converting $L$ into ${\cal R}$ and subtracting the Laughlin ground
state energy and the energy of two appropriate QP's from the total 
$N$-electron energy, $V_{AB}({\cal R})=E(L)-E_0-\varepsilon_A-
\varepsilon_B$.
To minimize the finite-size effects, all subtracted energies are given 
in the same units of $e^2/\lambda_0$, where $\lambda_0=R/\sqrt{S_0}$ 
corresponds to $2S_0=3(N-1)$, i.e., to $\nu={1\over3}$.
The result for $V_{\rm QER-QER}$ and $N\le9$ is shown in 
Fig.~\ref{fig3}(b).
Clearly, obtained values of $V_{\rm QER-QER}({\cal R})$ still depend 
on $N$ and, for example, the positive sign characteristic of repulsion 
between equally charged particles is only restored in the $N^{-1}
\rightarrow0$ limit with $V_{\rm QER-QER}(1)$ of the order of 0.01~$e^2
/\lambda$ (compare with discussion of the signs of $V_{\rm QE-QE}$ and 
$V_{\rm QH-QH}$ in Ref.~\onlinecite{wojs-qp}).
However, it seems that the monotonic character of $V_{\rm QER-QER}
({\cal R})$ is independent of $N$.
More importantly, $V_{\rm QER-QER}$ is also a super-linear function 
of $L(L+1)$.
This implies\cite{wojs-pmb,quinn,wojs-five} Laughlin correlations 
and incompressibility at $\nu_{\rm QER}=(2p+1)^{-1}$, in analogy 
to the spin-polarized Laughlin states of QE's or QH's in Haldane's 
hierarchy picture.\cite{haldane1,wojs-hierarchy} 
The most prominent of QE$_{\rm R}$ Laughlin states, $\nu_{\rm QER}
={1\over3}$, corresponds to the electronic filling factor of $\nu=
{4\over11}$ and the 75\% spin polarization ($J={1\over4}N$).
Since the $\nu_{\rm QE}={1\over3}$ state is compressible,
\cite{wojs-hierarchy} the experimental observation\cite{stormer} 
of the FQHE at $\nu={4\over11}$ seems to prove the formation of 
QE$_{\rm R}$'s in the $\nu={1\over3}$ state without need for direct 
measurement of spin polarization.
The expected critical dependence of the excitation gap at 
$\nu={4\over11}$ on the Zeeman gap $E_{\rm Z}$ might be revealed 
in tilted-field experiments.
This dependence will be very different than at some other fractions.
For example, the fact that incompressibility at $\nu={2\over5}$ can 
be a result of either maximally spin-polarized $\nu_{\rm QE}=1$ or 
completely spin-unpolarized ($J=0$) $\nu_{\rm QER}=1$ state gives rise 
to FQHE at this filling in both small and large $E_{\rm Z}$ regime.
On the other hand, spin-unpolarized FQHE is not expected in the
${1\over4}<\nu<{1\over3}$ range (because spin-reversed QH's in the
$\nu={1\over3}$ state do not exist), and the $\nu={2\over7}$ and
${4\over13}$ states (corresponding\cite{wojs-hierarchy} to 
$\nu_{\rm QH}={1\over3}$ and ${1\over5}$) should remain incompressible 
and compressible, respectively, over a wide range of $E_{\rm Z}$.

\begin{figure}[t]
\epsfxsize=3.40in
\epsffile{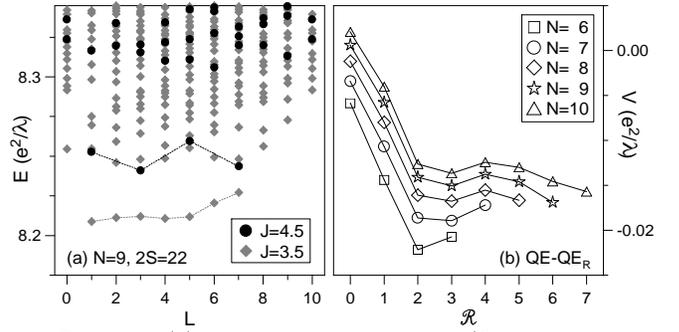}
\caption{
   (a) The energy spectrum (Coulomb energy $E$ versus angular 
   momentum $L$) of the system of $N=9$ electrons on Haldane 
   sphere at the monopole strength $2S=3(N-1)-2=22$.
   Black dots and grey diamonds mark states with the total 
   spin $J={1\over2}N={9\over2}$ (maximum polarization) and 
   $J={1\over2}N-1={7\over2}$ (one reversed spin), respectively.
   Lines connect states containing one QE--QE ($J={9\over2}$)
   or QE--QE$_{\rm R}$ ($J={7\over2}$) pair.
   (b) The pseudopotentials (pair energy $V$ versus relative angular 
   momentum ${\cal R}$) of the QE--QE$_{\rm R}$ interaction
   calculated in the systems of $N\le10$ electrons on Haldane sphere.
   $\lambda$ is the magnetic length.}
\label{fig4}
\end{figure}

The QE--QE$_{\rm R}$ pseudopotentials were calculated from similar
spectra as that of $J=3$ in Fig.~\ref{fig3}(a).
As another example, in Fig.~\ref{fig4}(a) we show the spectrum for
$N=9$, in which only two values of $J={1\over2}N={9\over2}$ and 
${1\over2}N-1={7\over2}$ have been included.
The lowest energy states in these two $J$-subspaces (connected with 
dashed lines) contain a QE--QE and QE--QE$_{\rm R}$ pair, respectively.
Using the same procedure as for $V_{\rm QER-QER}$, we have calculated
$V_{\rm QE-QER}({\cal R})$.
The results for $N\le10$ are presented in Fig.~\ref{fig4}(b).
As for $V_{\rm QE-QER}$ in Fig.~\ref{fig3}(b), the values of 
$V_{\rm QE-QER}({\cal R})$ calculated in a finite system depend 
on $N$.
The values extrapolated to the $N^{-1}\rightarrow0$ limit are 
also similar, with $V_{\rm QE-QER}(0)\rightarrow0.015\,e^2/\lambda$ 
and $V_{\rm QE-QER}(1)\rightarrow0.01\,e^2/\lambda$.

Despite finite-size errors, the comparison of the curves for $N\le10$ 
is sufficient to notice quite different behavior of $V_{\rm QE-QER}
({\cal R})$ from both $V_{\rm QER-QER}({\cal R})$ and $V_{\rm QE-QE}
({\cal R})$.
Two important features of the $V_{\rm QE-QER}$ pseudopotential can be
established: (i) the QE--QE$_{\rm R}$ repulsion is relatively strong 
at ${\cal R}\le1$ (short range) and saturates at larger ${\cal R}$, 
and (ii) $V_{\rm QE-QER}$ is super-linear in $L(L+1)$ only at $1\le
{\cal R}\le3$, but sub-linear at $0\le{\cal R}\le2$ and at larger 
${\cal R}$.
As a consequence, the short-range criterion\cite{wojs-pmb,quinn,%
wojs-five} applied to $V_{\rm QE-QER}$ yields Laughlin correlations 
for QE--QE$_{\rm R}$ pairs only at $m=2$.
The term ``Laughlin correlations'' used here is generally defined
\cite{haldane3,wojs-pmb,wojs-five} as a tendency to avoid pair 
states with ${\cal R}$ smaller than certain $m$.
At $\nu\le m^{-1}$, these correlations are described by a Jastrow 
prefactor $\prod_{ij}(x_i-y_j)^m$ in the many-body wave function 
($x$ and $y$ are complex QE and QE$_{\rm R}$ coordinates, 
respectively).

Although it is not clear if QE's and QE$_{\rm R}$'s could coexist in 
the $\nu={1\over3}$ ``parent'' state in an experimental system (such
mixed state would be sensitive to the value of $E_{\rm Z}$), one can 
ask if such two-component QE--QE$_{\rm R}$ plasma could also be 
incompressible.
This question can be answered within the generalized CF model
\cite{quinn,wojs-cf} for all allowed combinations of Jastrow exponents 
[$m_{\rm QE-QE},m_{\rm QER-QER},m_{\rm QE-QER}$].
In this model, the reduced (effective) LL degeneracies of QP's are 
given by $g_{\rm QE}^*=g_{\rm QE}-(m_{\rm QE-QE}-1)(N_{\rm QE}-1)-
m_{\rm QE-QER}N_{\rm QER}$ and $g_{\rm QER}^*=g_{\rm QER}-(m_{\rm 
QER-QER}-1)(N_{\rm QER}-1)-m_{\rm QE-QER}N_{\rm QE}$, and the 
incompressibility condition is $N_{\rm QP}=g_{\rm QP}^*$ for both 
QE's and QE$_{\rm R}$'s.
In the above, $g$ is the LL degeneracy of electrons and $N_{\rm QP}$ 
denotes the number of QP's of each type.
It turns out that because the three involved QP pseudopotentials 
are not generally super-linear in $L(L+1)$, only few combinations 
of exponents [$m_{\rm QE-QE},m_{\rm QER-QER},m_{\rm QE-QER}$] are 
allowed, and of those only [1,1,2] satisfies the incompressibility 
condition.
The hypothetical [1,1,2] state of the QE--QE$_{\rm R}$ fluid corresponds
to $\nu={5\over13}$ and 80\% polarization ($J={3\over10}N$).
Finite realizations of this state on Haldane's sphere occur for 
$N=5q+4$ ($q\ge1$) at $2S=13q+7$, and have $N_{\rm QE}=q$ 
and $N_{\rm QER}=q+2$, which yields $J={3\over2}q$.

\subsection{Optical Properties}

Once the single-particle energies $\varepsilon$ and the two-particle 
interaction pseudopotentials $V({\cal R})$ of all three types of QP's 
have been calculated, let us now turn to their optical properties.
The effect of QE's on the photoluminescence (PL) spectrum of the 
Laughlin fluid has been studied in great detail.\cite{quinn,wojs-fcx}
The crucial facts are:
(i) The PL spectrum can be understood in terms of QE's and their
interaction with one another and with a valence-band hole ($h$) only in 
the ``weak-coupling regime'' in which the electron--electron repulsion 
is sufficiently weak compared to the electron--hole attraction; this is 
realized in ``asymmetric'' structures in which the electron and hole 
layers are separated by a finite distance $d$ (of the order of $\lambda$).
(ii) In this regime, a positively charged $h$ can bind one or two QE's
to form ``fractionally charged excitons'' (FCX), $h$QE or $h$QE$_2$.
(iii) The 2D translational invariance results in orbital selection rules
for the radiative recombination of FCX's; it turns out that the only 
bright states are $h$QE* (an excited state of the dark $h$QE) and 
$h$QE$_2$.

\begin{figure}[t]
\epsfxsize=3.40in
\epsffile{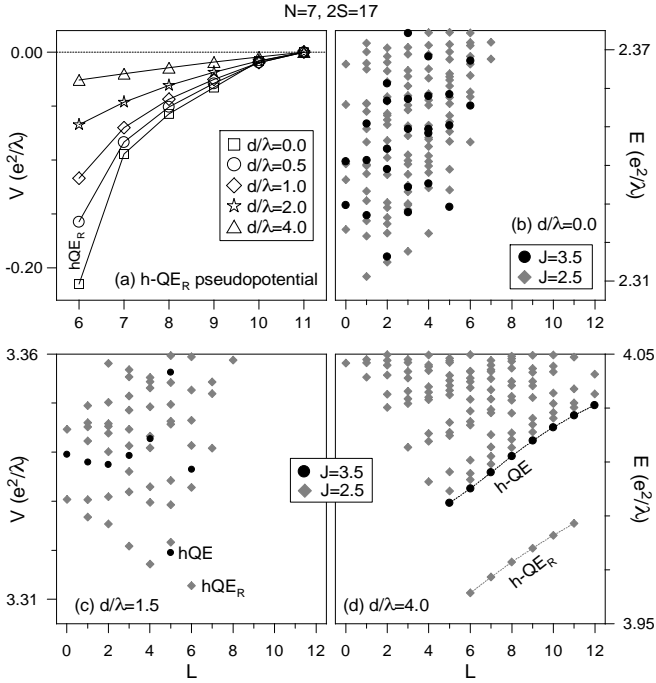}
\caption{
   (a) The pseudopotentials (pair energy $V$ versus pair angular 
   momentum $L$) of the $h$--QE$_{\rm R}$ interaction calculated 
   in the system of $N=7$ electrons and one valence hole ($h$) 
   on Haldane sphere at the monopole strength $2S=3(N-1)-1=17$.
   Different symbols correspond to different separations $d$ 
   between the electron and hole planes.
   (bcd) The energy spectra (Coulomb energy $E$ versus angular 
   momentum $L$) of the same, seven-electron--one-hole system
   at $2S=17$ at three different values of $d$.
   Black dots and grey diamonds mark states with the total 
   electron spin $J={1\over2}N={7\over2}$ (maximum polarization) 
   and $J={1\over2}N-1={5\over2}$ (one reversed spin), 
   respectively.
   Lines in (d) connect states containing one $h$--QE 
   ($J={7\over2}$) or $h$--QE$_{\rm R}$ ($J={5\over2}$) pair.
   The lowest-energy $J={7\over2}$ and ${5\over2}$ states 
   in (c) are the fractionally charged excitons, 
   $h$QE and $h$QE$_{\rm R}$, respectively.
   $\lambda$ is the magnetic length.}
\label{fig5}
\end{figure}

In analogy, we expect that a valence hole $h$ could also form bound 
states with one or more QE$_{\rm R}$'s, denoted by FCX$_{\rm R}$.
However, unlike for FCX's, the stability of FCX$_{\rm R}$ complexes 
should depend on the Zeeman energy, the binding of more than one 
QE$_{\rm R}$ should be more difficult due to the stronger 
QE$_{\rm R}$--QE$_{\rm R}$ repulsion, different angular momenta of 
QE and QE$_{\rm R}$ should result in different optical selection 
rules of FCX$_{\rm R}$, and the possible annihilation of a hole with 
a reversed-spin electron should cause different polarization of 
FCX$_{\rm R}$ emission.
To study the possible binding of FCX$_{\rm R}$'s we begin with the
$h$--QE$_{\rm R}$ pseudopotential, shown in Fig.~\ref{fig5}(a) for 
a $7e$--$h$ system in which a hole interacts with $N=7$ electrons 
and for a few different values of $d/\lambda$.
The values of $2S=3(N-1)-1=17$ and $J={1\over2}N-1={5\over2}$
are chosen so that one QE$_{\rm R}$ is present in the Laughlin
$\nu={1\over3}$ state and interacts with the hole.
In the CF picture of this configuration, $2S^*=2S-2(N-1)=5$ so that 
six CF's fill completely the lowest CF LL of $g^*=2S^*+1$, leaving
the seventh CF in the reversed-spin LL.
The filled LL is incompressible, and only the single reversed-spin 
CF (i.e., QE$_{\rm R}$) correlates with the hole.
The $V_{h{\rm -QER}}$ is plotted as a function of the pair angular
momentum whose values ($6\le L\le11$) result from addition of 
$l_h=S$ and $l_{\rm QER}=S^*$.
To ensure that exactly one QE$_{\rm R}$ is present in the Laughlin 
fluid and interacts with the hole at an arbitrary (small) value of 
$d$, a special procedure\cite{wojs-fcx} has been used in which the 
electric charge of the hole is reduced to $e/\epsilon\ll e$.
Clearly, the decrease of $V_{h{\rm -QER}}$ with a decrease of $L$
(average $h$--QE$_{\rm R}$ separation) indicates $h$--QE$_{\rm R}$
attraction.
The strength of this attraction, that is the binding energy 
$\Delta_{h{\rm QER}}\sim|V_{h{\rm -QER}}(l_h-l_{\rm QER})|$,
depends on $d$ and is similar to $\Delta_{h{\rm QE}}$; compare 
with Ref.~\onlinecite{wojs-fcx}.
Therefore, in analogy to the QE case, we expect that bound 
$h$QE$_{\rm R}$ states will occur in a system containing free 
QE$_{\rm R}$'s at the values of $d$ at which $\Delta_{h{\rm QE}}$
and $\Delta_{h{\rm QER}}$ is smaller than the Laughlin gap to create
additional QE--QH pairs (note that since the projection $J_z$ of the 
total electron spin is conserved at any $d$, FCX or FCX$_{\rm R}$
does not couple to virtual QE$_{\rm R}$--QH excitations).

In order to verify the above hypothesis, we have calculated the 
$7e$--$h$ energy spectra with up to one reversed spin ($J={1\over2}N
={7\over2}$ and $J={1\over2}N-1={5\over2}$).
The results for $d/\lambda=0$, 1.5, and 4 are presented in 
Fig.~\ref{fig5}(bcd).
As expected, the $h$QE$_{\rm R}$ ground state develops together with
the spin-polarized $h$QE state at $d$ larger than about $\lambda$.
The energy difference between $h$QE$_{\rm R}$ and $h$QE states at 
$d/\lambda=1.5$ is only about 0.007~$e^2/\lambda$, which is small 
compared to $\varepsilon_{\rm QE}-\varepsilon_{\rm QER}$.
This is because $h$QE couples stronger than $h$QE$_{\rm R}$ to 
virtual QE--QH pair excitations of the underlying Laughlin state 
(QE--QE$_{\rm R}$ repulsion at short range is stronger than QE--QE 
repulsion).
At $d$ much larger than $\lambda$, the lowest energy states in 
Fig.~\ref{fig5}(d) contain well defined $h$--QE or $h$--QE$_{\rm R}$ 
pairs with all possible values of $L$.
The coupling to the virtual QE--QH excitations is reduced, and the
$h$--QE$_{\rm R}$ and $h$--QE bands are separated by about the
single-particle gap $\varepsilon_{\rm QE}-\varepsilon_{\rm QER}$.

\begin{figure}[t]
\epsfxsize=3.40in
\epsffile{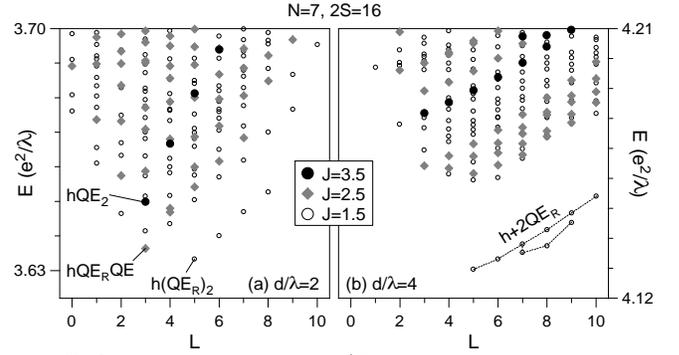}
\caption{
   The energy spectra (Coulomb energy $E$ versus angular momentum 
   $L$) of the system of $N=7$ electrons and one valence hole ($h$) 
   on Haldane sphere at the monopole strength $2S=3(N-1)-2=16$, 
   at the separations $d=2\lambda$ (a) and $4\lambda$ (b) between 
   the electron and hole planes.
   Black dots, grey diamonds, and open circles mark states with the 
   total electron spin $J={1\over2}N={7\over2}$ (maximum polarization),
   $J={1\over2}N-1={5\over2}$ (one reversed spin), and
   $J={1\over2}N-2={3\over2}$ (two reversed spins), respectively.
   The lowest-energy $J={7\over2}$, ${5\over2}$, and ${3\over2}$ 
   states in (a) are the fractionally charged excitons, $h$QE$_2$, 
   $h$QE$_{\rm R}$QE, and $h($QE$_{\rm R})_2$, respectively.
   The lowest-energy band of $J={3\over2}$ states marked with lines 
   in (b) contains all possible states of two QE$_{\rm R}$'s and 
   one $h$.
   $\lambda$ is the magnetic length.}
\label{fig6}
\end{figure}

To compare the optical properties of $h$QE and $h$QE$_{\rm R}$,
it is essential to notice that, because $l_{\rm QER}\ne l_{\rm QE}$, 
also $l_{h{\rm QER}}=l_h-l_{\rm QER}=N-1$ is different from
$l_{h{\rm QE}}=l_h-l_{\rm QE}=N-2$.
The orbital selection rule for radiative recombination of bound FCX
or FCX$_{\rm R}$ states results from the fact that an annihilated,
optically active electron--hole pair carries no angular momentum.
\cite{wojs-fcx,wojs-xminus,avron,dzyubenko}
Therefore, the angular momenta of the initial (bound) state and
a final state in the emission process must be equal.
On the other hand, it is known\cite{chen,wojs-fcx} that only those 
emission processes with minimum number of QP's involved can have 
significant spatial overlap with an initial (bound) state of small 
size, and thus significant intensity (oscillator strength $\tau^{-1}$).
Thus, $h$QE or $h$QE$_{\rm R}$ must both recombine to leave two QH's 
in the final state (and no additional QE--QH or QE$_{\rm R}$--QH pairs).
The allowed angular momenta of two identical QH's [in the final, 
$(N-1)e$ system] each with $l_{\rm QH}={1\over2}N$ are $L_{\rm 2QH}
=N-{\cal R}_{\rm QH}$, where ${\cal R}_{\rm QH}$ is an odd integer.
The comparison of $L_{\rm 2QH}$ with $l_{h{\rm QE}}$ and $l_{h{\rm 
QER}}$ makes it clear that, in contrast to the dark $h$QE, the 
$h$QE$_{\rm R}$ ground state is radiative.
Since $h$QE$_{\rm R}$ is the simplest of all FCX$_{\rm R}$'s and 
bright at the same time, its emission is expected to dominate the 
PL spectrum of a Laughlin fluid at $\nu>{1\over3}$, in which free
QE$_{\rm R}$'s are present.
The larger FCX$_{\rm R}$ complexes, $h$(QE$_{\rm R}$)$_2$ and
$h$QE$_{\rm R}$QE are also found in the numerical calculation at 
$d>\lambda$ (see Fig.~\ref{fig6}), but being less strongly bound
(due to larger QE$_{\rm R}$--QE$_{\rm R}$ and QE$_{\rm R}$--QE 
repulsion at short range) they are not expected to form as easily
as $h$QE$_2$ does in a spin-polarized system.
Moreover, $h$(QE$_{\rm R}$)$_2$ turns out dark, and the formation 
of $h$QE$_{\rm R}$QE requires the presence of both QE's and 
QE$_{\rm R}$'s in the unperturbed electron system, which further
limits the contribution of these bound states to the PL spectrum.
Let also add that since $h$QE$_{\rm R}$ emits by recombination of 
a valence hole with ${1\over3}$ of an electron with reversed spin 
(QE$_{\rm R}$ in the initial state) and ${2\over3}$ of an electron 
with majority spin (two QH's in the final state), the emitted photon 
should be only partially polarized.
This is in contrast to a completely polarized emission of the bright
FCX complexes, $h$QE* and $h$QE$_2$.
Therefore, the partially unpolarized emission in the ``weak-coupling'' 
regime ($d>\lambda$) could be an indication of the presence of 
QE$_{\rm R}$'s in the electron fluid.

\section{Conclusion}
Using exact numerical diagonalization, we have studied the low-energy 
spin-flip excitations of a 2DEG in the FQH regime (at $\nu={1\over3}$), 
so-called reversed-spin quasielectrons (QE$_{\rm R}$'s).
The pseudopotentials $V({\cal R})$ describing interaction of 
QE$_{\rm R}$'s with one another and with other Laughlin QP's have 
been calculated.
From the form of the QE$_{\rm R}$--QE$_{\rm R}$ pseudopotential it 
is shown that the Haldane-hierarchy $\nu={1\over3}$ daughter state 
of QE$_{\rm R}$'s formed in the parent $\nu={1\over3}$ Laughlin state 
of electrons is incompressible.
This state corresponds to the total electron filling factor of 
$\nu={4\over11}$ and partial, 75\% spin polarization.
Because the analogous $\nu={1\over3}$ hierarchy state of QE's is 
known to be compressible, it is claimed that the experimentally 
observed\cite{stormer} FQHE at $\nu={4\over11}$ confirms the formation 
of QE$_{\rm R}$'s and their Laughlin correlations in a 2DEG with low 
Zeeman splitting.
Although the stability of mixed QE--QE$_{\rm R}$ hierarchy states is 
expected to be highly sensitive to the Zeeman energy $E_{\rm Z}$, it 
is predicted that an incompressible [1,1,2] state that corresponds to 
$\nu={5\over13}$ and 80\% spin polarization might form at appropriate 
$E_{\rm Z}$.
The interaction of QE$_{\rm R}$'s with a spatially separated 
valence-band hole has also been studied.
In analogy to the so-called fractionally charged exciton (FCX) states 
$h$QE$_n$, the spin-reversed complexes FCX$_{\rm R}$ that involve 
one or more QE$_{\rm R}$'s are predicted.
Because QE and QE$_{\rm R}$ have different angular momenta, the 
optical selection rules for FCX and FCX$_{\rm R}$ are different, 
and, for example, $h$QE$_{\rm R}$ turns out radiative in contrast 
to the dark $h$QE, while $h($QE$_{\rm R})_2$ is dark in contrast 
to the bright $h$QE$_2$, 
Therefore, in addition to obvious difference in polarization, the 
emission from FCX and FCX$_{\rm R}$ states is expected to occur at 
a different energy and differently depend on temperature.

\section{Acknowledgment}
The authors acknowledge partial support by the Materials Research 
Program of Basic Energy Sciences, US Department of Energy.
AW thanks L. Jacak, M. Potemski, and P. Hawrylak for discussions.
AW and IS acknowledge support from grant 2P03B11118 of the Polish 
State Committee for Scientific Research (KBN).



\end{document}